\documentclass[12pt]{iopart}

\usepackage{amssymb}
\usepackage{color}

\usepackage{graphicx,epsfig}

\newcommand{\<}{\langle}
\renewcommand{\>}{\rangle}

\begin{document}

\title{Entanglement at the quantum phase transition in a harmonic lattice}

\author{Elisabeth Rieper$^1$, Janet Anders$^{2}$ and Vlatko Vedral$^{1,3,4}$}

\address{$^1$ Center for Quantum Technologies, National University of Singapore, Republic of Singapore}
\address{$^2$ Department of Physics and Astronomy, University College London, London WC1E 6BT, United Kingdom}
\address{$^3$ Atomic and Laser Physics, Clarendon Laboratory, University of Oxford, Parks Road, Oxford OX13PU, United Kingdom}
\address{$^4$ Department of Physics, National University of Singapore, Republic of Singapore}

\eads{ \mailto{ER: elisabeth.rieper@quantumlah.org}}

\begin{abstract}
The entanglement properties of the phase transition in a two dimensional harmonic lattice, similar to the one observed in recent ion trap experiments, are discussed both, for finite number of particles and thermodynamical limit. We show that for the ground state at the critical value of the trapping potential two entanglement measures, the negativity between two neighbouring sites and the block entropy for blocks of size 1, 2 and 3, change abruptly. Entanglement thus indicates quantum phase transitions in general; not only in the finite dimensional case considered in \cite{Lidar}. Finally, we consider the thermal state and compare its exact entanglement with a temperature entanglement witness introduced in \cite{Anders08}.
\end{abstract}



\maketitle

\section{Introduction}
Coupled harmonic chains with short and long range interactions are ubiquitous in science and engineering. Their application to calculate the phononic heat capacity by Einstein \cite{Einstein} marks the birth of solid state physics. Beyond physics, harmonic chains feature in chemistry and biology, where they are used to model behaviour of macro-molecules, such as DNA \cite{DNAHO} and cell membranes \cite{Froehlich}. In the last decade harmonic systems have been revised using techniques developed in quantum information science to study correlation properties in the quantum regime and particularly at small temperatures \cite{Audenaert,Paz,Eisert,Galve}. Thermodynamics has been very successful in characterising ``standard'' phase transitions that occur at finite temperature when a macroscopic parameter, such as pressure, is changed \cite{Yeomans}. Quantum phase transitions (QPTs) appear at zero temperature \cite{Sachdev} and are due to the change of an external parameter, such as the trapping potential of an ion trap. These transitions are driven by quantum fluctuations and have been linked to entanglement for the case of finite dimensional systems \cite{Fazio,Lidar}. 

In this paper we study a QPT in a continuous variable system: a system of trapped ions which we model as a harmonic lattice \cite{EisertPlenio,AndersDipl}. The ions interact via a long-range Coulomb repulsion and are trapped by two external  potentials, see Fig. \ref{fig:trap}. They align in a linear configuration for big enough transversal trapping potential, $\nu_t$ however when $\nu_t$ is decreased the system undergoes a phase transition and the new equilibrium state forms a zig-zag configuration. This model is motivated by ion trap experiments \cite{1stExp,iontrap,Blatt}, where such a QPT occurs \cite{firstExp,ExpPowerLaw}. Recent analytical studies of the transition using Landau theory \cite{Classical} allowed to determine the system's classical behaviour at the transition point. Moreover, the numerical treatment of the quantised system of a few ions promised the possibility of simulating linear and nonlinear Klein-Gordon fields on a lattice \cite{Plenio}. However, a comprehensive analytical study of the quantised system has so far been lacking due to the complexity of the system. 

Here we model the ion trap scenario as a lattice of harmonically coupled oscillators and present a first quantitative characterization of the entanglement inherent in both ground state configurations of the ions. For finite dimensional systems QPTs of first (second) order are characterised by a discontinuity (a discontinuity in or divergence of the first derivative) of the negativity \cite{Lidar}. We show that also in the here considered  continuous variable system the structure of entanglement, measured by the negativity and the von Neumann entropy, changes abruptly at the critical point and indicates the occurrence of a QPT, in a similar way as classical correlations indicate standard phase transitions. The first derivative of the negativity between two neighbouring ions has a finite discontinuity and the von Neumann entropy of contiguous blocks of a single ion, two and three ions all show a divergence in the first derivative. The long-ranged nature of the Coulomb interaction leads to an increase of the block entropy with increasing block size. This is in contrast to models with only nearest neighbour interaction where `area laws' apply \cite{ReviewAreaLaw} and entanglement does not increase with block size, i.e. volume, as long as the surface of the block is constant. However, our results show that the increase in block entropy with the block size is quiet small due to the fast decline of the Coulomb potential. 

The quantum fluctuations that cause the QPT are most dominant at zero temperature. However, as experiments are performed at small, but finite temperatures it is important to know how temperature affects these fluctuations \cite{Kopp05}. In the final part of our paper we discuss thermal states and find that the sharpness of the QPT, indicated by the entanglement, fades out with increasing temperature. Another macroscopic consequence of quantum fluctuations is the lowering of the energy of the system \cite{Anders08}. We compute up to which temperature the thermal state has a lower energy than any separable state. This temperature witness could be implemented in an ion trap experiment by measuring the average energy (i.e. mean excitation). Although the model is motivated by ion traps realisations include the vibrational motion of molecules \cite{DNAHO} (nuclei in the electronic potential) and optical lattices \cite{Bloch}.

\section{The model}
\begin{figure}[t]
	\begin{center}
	\includegraphics[width=0.7\textwidth]{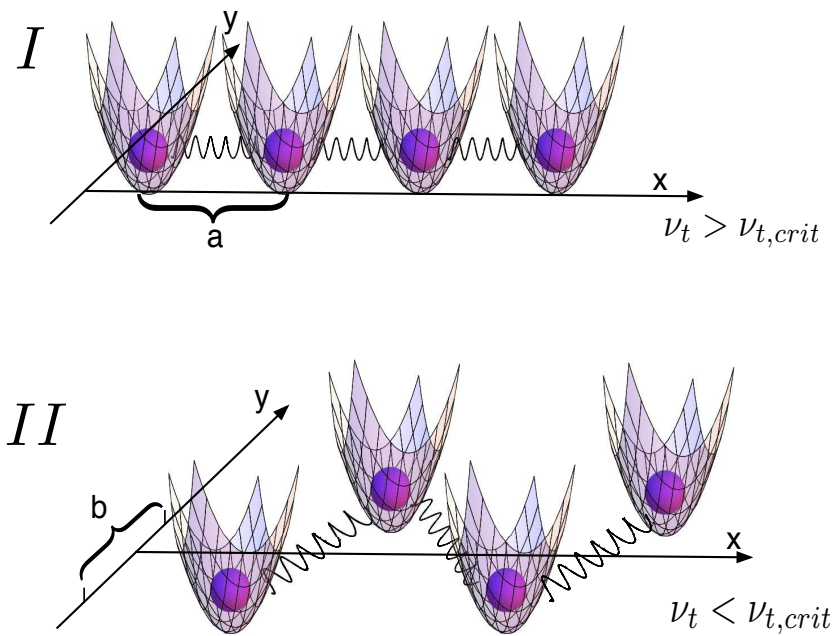}
	\caption{\label{fig:trap} Sketch of the harmonic lattice under consideration: Each site is trapped by two external potentials, $\nu$ in $x$ direction and $\nu_t$ in $y$ direction. For clearness only the nearest neighbour coupling is indicated, however all ions interact via a long-range potential which we approximate harmonically. This could be for instance a Coulomb potential such as is common in ion experiments. The equilibrium distances between the sites are assumed to be equidistant, with lattice constant $a$ in $x$-direction and $b$ along the $y$-direction. In $I$, the transverse trapping potential $\nu_t$ is larger than the critical value and the ions arrange linearly. As displayed in $II$, decreasing  the transverse trapping potential below the critical value $\nu_{t, crit}$ leads to a QPT causing the ions to move outwards and form a two-dimensional zig-zag structure. 
	}
	\end{center}
\end{figure}

We consider the Hamiltonian, see Fig. \ref{fig:trap}, 
\begin{equation}
	H (\nu_t) = \sum_{j=1}^N \left(\frac{p_{x_j}^2+p_{y_j}^2}{2m}+\frac{m \nu^2 }{2}(\tilde{x}_j-\tilde{x}_{j,0})^2
	+\frac{m \nu_t^2}{2} \tilde{y}_j^2 +V_j\right),
\end{equation}
where $V_j=\frac{1}{2}\sum_{k \neq j} \frac{Q^2}{|\vec{r}_k-\vec{r}_j|}$ are the Coulomb potentials of the sites with $\vec{r}_j=(\tilde{x}_j,\tilde{y}_j)$ absolute coordinates of the sites and $\tilde{x}_{j,0}$ are the equilibrium positions in $x$ direction. $N$ is the number of particles, $Q$ is the charge of the ions and  $\nu$ ($\nu_t$) are the trapping potentials in $x$ ($y$) direction. We assume periodic boundary conditions, $\vec{r}_j=\vec{r}_{j+N}$. 

To calculate the entanglement measures we approximate the Coulomb potential to second order and expand about the equilibrium positions. The key step is then to diagonalise the Hamiltonian into a set of uncoupled modes, the lattice vibrations, with which analytic expressions for the measures can be obtained. Similar to the classical calculation \cite{Classical}, we use a simplified model with equidistant equilibrium position in $x$ direction, spaced by the lattice constant $a$. Such condition can be realised for the central ions of a long ion chain inside  a linear Paul trap \cite{lintrap} or for ions confined in a ring of large radius \cite{firstExp,circletrap}. 

For big trapping potential the sites are arranged on a single line, i.e. $\tilde{x}_{j,0}= a \, j$ and $\tilde{y}_{j,0}=0$, while for small enough $\nu_t$, the equilibrium positions become $\tilde{x}_{j,0}= a \, j$ and $\tilde{y}_{j,0}=(-1)^j \frac{b}{2}$ and a two-dimensional zig-zag configuration is formed. The equation determining the deviation $b$ in $y$ direction is obtained by summing the linear terms over all sites and requiring it to vanish,
\begin{equation}
	\frac{1}{2} m \nu_t^2=Q^2 \sum_{\tau=2l+1} \frac{1}{\sqrt{\tau^2 a^2+b^2}^3},
\end{equation}
where $\tau=k-j$ numbers the neighbours of each sites. The harmonically approximated Hamiltonian becomes
\begin{eqnarray} \label{eq:Hami}
	H=\sum_{j=1}^N \left( \frac{p_{xj}^2+p_{yj}^2}{2m} + \frac{m \nu^2}{2}x_j^2+\frac{m \nu_t^2}{2} y_j^2 \right)  \\ \nonumber
	+\frac{Q^2}{2} \sum_{j=1}^N \sum_{\tau > 0} \left(
		d_\tau^x \, (x_{j}-x_{j+\tau})^2 + d_\tau^y \, (y_{j}-y_{j+\tau})^2 
			+ d_\tau^{xy} \, (x_{j}-x_{j+\tau})(y_{j}-y_{j+\tau}) \right),
\end{eqnarray}
with $x_j=\tilde{x}_j-\tilde{x}_{j,0}$ and $y_j=\tilde{y}_j-\tilde{y}_{j,0}$ the deviations from equilibrium. Furthermore, the $d_{x, y, xy}$ denote the second order Taylor coefficients of the Coulomb potential which are, for the linear and zig-zag configuration,
\begin{eqnarray}
	d_\tau^x=\frac{1}{ (a \, \tau)^3} 
		&\mbox{ and }&	
	d_\tau^x= \frac{2\tau^2 a^2-\delta_{\tau, odd} \, b^2}{  2\sqrt{(\tau a )^2+ \delta_{\tau, odd} \, b^2}^5 },\\
	d_\tau^y= - \frac{1}{2 (a \tau)^3}
		&\mbox{ and }&	
	d_\tau^y=\frac{2\delta_{\tau, odd} \, b^2-\tau^2 a^2}{2\sqrt{(\tau a )^2+ \delta_{\tau, odd} \, b^2}^5}, \\
	d_\tau^{xy}= 0
		&\mbox{ and }&	
	d_\tau^{xy}= \delta_{\tau, odd} \, (-1)^j \frac{3 \tau a  b}{2\sqrt{(\tau a )^2+b^2}^5}.
\end{eqnarray}

\section{Calculation of entanglement measures}
We are interested in the behaviour of the entanglement between the sites in the chain for varying transverse trapping potential, $\nu_t$, and particularly at the point of criticality, $\nu_{t, crit}$. We calculate the  \emph{negativity}, $E_N$, for two modes regardless of all others, for instance, the $x$ degrees of freedom of two (neighbouring) sites. To measure the correlation of one mode with all other modes, for instance the entanglement between the $y$ degree of freedom of a  single site or block of sites with all the other degrees of freedom in the chain, the \emph{von Neumann entropy}, $S_V$, is used. 
 Other entanglement measures are available, such as the entanglement of formation \cite{EntofForm}. Yet they are very hard to calculate in this continuous variable scenario and we will be content with the two measures as stated. To see the effect of the long-range interaction we compare the full long-range (LR) Coulomb Hamiltonian with a cut-off version in which only nearest neighbours interact (NN) and the interaction with more distant neighbours is set to zero.
 
For the \emph{linear configuration} the Hamiltonian decouples into $x$ and $y$ part. A discrete Fourier transformation for the $x$ (similar in $y$ direction) of the form
\begin{eqnarray} \label{eq:FTlinear}
	x_j=\frac{1}{\sqrt{N}}\sum_{l=1}^N e^{i\frac{2\pi}{N}jl} X_l 
		\, \mbox{ and } \,
	p_{xj}=\frac{1}{\sqrt{N}}\sum_{l=1}^N e^{-i\frac{2\pi}{N}jl} P_{xl} \hbox{ ,}
\end{eqnarray}
maps the space coordinates of the sites into diagonal modes, the lattice vibrations or phonons. The diagonal Hamiltonian is then $H=\sum_{l=1}^N \hbar \omega_{xl} \left(\hat{n}_{xl}+\frac{1}{2}\right)+\sum_{l=1}^N \hbar \omega_{yl} \left(\hat{n}_{yl}+\frac{1}{2}\right)$ where $\hat{n}_{xl} = \frac{P_{xl}^{\dag} \, P_{xl}}{2 \hbar \, m \omega_{xl}} + \frac{m \omega_{xl} \, X_l^{\dag} \, X_l }{2 \hbar} - \frac{1}{2}$, and similarly $\hat{n}_{yl}$, are the number operators in $x$ and $y$ direction for mode $l$, and the frequencies are 
\begin{eqnarray}
	\omega_{xl} = \sqrt{\nu^2 + C \sum_{\tau > 0} \frac{\sin^2\left(\frac{\pi l \tau}{N}\right)}{\tau ^3}}
			\, \mbox{ and } \,
	\omega_{yl} = \sqrt{\nu_t^2-\frac{C}{2} \sum_{\tau > 0}\frac{\sin^2\left(\frac{\pi l \tau}{N}\right)}{\tau ^3}}
\end{eqnarray} 
where $C=\frac{4 Q^2}{m a^3}$. The asymmetry of the system is reflected in the dispersion relations. $\omega_{xl}$ is always real, whereas $\omega_{yl}$ would become complex for small values of transverse trapping potential $\nu_t$. This is where the quantum phase transition occurs. The critical value $\nu_{t,crit}$ for LR interaction is $\nu_{t,crit}\approx \sqrt{0.6 C}$ while for NN interaction it is $\nu_{t,crit}= \sqrt{0.5 C}$.

To diagonalise the Hamiltonian in the emerging \emph{zig-zag configuration}, where the previously independent phonons now couple in $x-y$-direction, see Eq.~(\ref{eq:Hami}), we need to amend the transformation in the $y$ direction Eq.~(\ref{eq:FTlinear}) and transform with  
\begin{eqnarray} \label{eq:trafo-zig-zag}
	y_j=\frac{i}{\sqrt{N}}\sum_l e^{i\frac{2\pi}{N}j(l+N/2)} Y_l 
		\, \mbox{ and } \,
	p_{yj}=\frac{i}{\sqrt{N}}\sum_l e^{-i\frac{2\pi}{N}j(l+N/2)} P_{yl}.
\end{eqnarray}
The additional factor $e^{i \pi j}=(-1)^j$ compensates the alternating sign of the $x-y$ coupling in the Hamiltonian. Expressed with a coupling matrix 
\begin{equation} M_l =  
	\left(
\begin{array}{cccc}
	\frac{m}{2} \tilde\omega_{xl}^2& 0 &\frac{m}{2} \tilde\omega_{xyl} & 0  \\
	0 & \frac{1}{2m} &   0 & 0  \\
	\frac{m}{2} \tilde\omega_{xyl} & 0 &\frac{m}{2}  \tilde\omega_{yl}^2 & 0  \\
	0 & 0 &   0 & \frac{1}{2m}  \\
\end{array}
	\right).
\end{equation}
the Hamiltonian can now be written as 
\begin{equation}
	H =  \sum_{l=1}^N (X_l^\dagger, P_{x_l}^\dagger, Y_l^\dagger,  P_{y_l}^\dagger) \, M_l \, \left( \begin{array}{c} X_l \\ P_{x_l} \\ Y_l \\ P_{y_l} \end{array} \right),
\end{equation}
with coefficients $\tilde\omega_{u,l}=\sqrt{\nu_{u}^2+\frac{4 Q^2}{m} \sum_{\tau >0} d_{u,\tau} \sin^2(\pi l \tau /N)}$, $u=x,y$ and $\tilde\omega_{xyl}= \frac{Q^2}{m} \sum_{\tau >0, odd} \frac{3 \tau a  b}{2\sqrt{(\tau a )^2+b^2}^5} \sin(2 \pi l \tau /N)$. 

For each $l$ we need the symplectic transformation \cite{Williamson}, denoted by  $S_l$, that diagonalises the matrix $M_l$,
\begin{equation}
	S_l \, M_l \, S_l^T = \mbox{diag} \left(\frac{\omega_{v, l}}{2}, \frac{\omega_{v, l}}{2}, \frac{\omega_{w, l}}{2}, \frac{\omega_{w, l}}{2} \right) 
\end{equation}
where 
\begin{eqnarray}
 	\omega_{v,l}=\frac{1}{2\sqrt{2}} 
		\sqrt{\tilde\omega_{xl}^2 + \tilde\omega_{yl}^2 +
			\sqrt{(\tilde\omega_{xl}^2-\tilde\omega_{yl}^2)^2+ 4\tilde\omega_{xyl}^2}}\\
	\omega_{w,l}=\frac{1}{2\sqrt{2}} 
		\sqrt{\tilde\omega_{xl}^2 + \tilde\omega_{yl}^2 -
			\sqrt{(\tilde\omega_{xl}^2-\tilde\omega_{yl}^2)^2+ 4\tilde\omega_{xyl}^2}}		
\end{eqnarray}
are the symplectic eigenvalues of $M_l$. These transformations are given by
\begin{equation} S_l =
\left(
\begin{array}{cccc}
	0& -\phi_l \sqrt{\frac{\omega_{v, l} m}{ \psi_l}} & 0 &-2 \tilde\omega_{xyl} \sqrt{\frac{\omega_{v, l}m}{ \psi_l}}  \\
	\frac{\phi_l}{\sqrt{2 m \omega_{v, l} \, \psi_l}} & 0 &   \frac{\tilde\omega_{xyl}}{\sqrt{\omega_{v, l}m \,  \psi_l}} & 0  \\
	0 &-2 \tilde\omega_{xyl} \sqrt{\frac{\omega_{v, l}m}{ \psi_l}} & 0 &\phi_l \sqrt{\frac{\omega_{v, l}m}{ \psi_l}} \\
 	\frac{\tilde\omega_{xyl}}{\sqrt{\omega_{v, l}m \, \psi_l}} & 0 &  -\frac{\phi_l}{2\sqrt{\omega_{v, l}m \, \psi_l}}  & 0 \\
\end{array}
\right)
\end{equation}
where $\phi_l = \tilde \omega_{xl}^2 - \tilde\omega_{yl}^2 + \sqrt{(\tilde\omega_{xl}^2 - \tilde\omega_{yl}^2)^2 + 4\tilde\omega_{xyl}^2}$ 
and $\psi_l = (\tilde\omega_{xl}^2 - \tilde\omega_{yl}^2)^2 + 4 \tilde\omega_{xyl}^2 + (  \tilde\omega_{xl}^2-\tilde\omega_{yl}^2) \sqrt{(\tilde\omega_{xl}^2 - \tilde\omega_{yl}^2)^2 +4 \tilde\omega_{xyl}^2}$.
The new normal modes are 
\begin{equation} \label{eq:diamodes}
	\left( \begin{array}{c} v_l \\ P_{v_l} \\ w_l \\ P_{w_l} \end{array} \right) = S_l^{-T} \, \left( \begin{array}{c} X_l \\ P_{x_l} \\ Y_l \\ P_{y_l} \end{array} \right),
\end{equation}
and using the number operators $\hat n_{v, l} = \frac{P^{\dagger}_{v_l} P_{v_l}}{2 \hbar } +  \frac{ v^{\dagger}_l v_l}{2\hbar}  - \frac{1}{2}$, and similarly for $w$, we find the fully diagonalised Hamiltonian for the zig-zag configuration $H = \sum_l \left[\hbar \omega_{v, l} \left(\hat n_{v, l} + {1\over 2}\right) + \hbar \omega_{w, l} \left(\hat n_{w, l} + {1\over 2}\right)\right]$.

The calculation of the two-site negativity requires the evaluation of the covariance matrix of the partially transposed state of the two sites \cite{EisertPlenio,AndersDipl}. In two dimensions this is a $8 \times 8$ matrix of which the symplectic eigenvalues  have to be found. However, in the linear configuration of the ions, $x$ and $y$ degree of freedom completely decouple and it is sufficient to consider two $4 \times 4$ covariance matrices independently. In contrast the zig-zag configuration contains $xy$-coupling terms and it is \emph{a priori} necessary to consider the full $8 \times 8$ matrix. As a result no entanglement occurs between the $x$ and $y$ direction in the linear configuration while the zig-zag configuration could sustain $xy$ entanglement. However, we found that all expectation values coupling the $x$ and $y$ direction, i.e. $\langle x_i \, y_j \rangle $, $\langle p_{x_i} \, p_{y_j} \rangle $ etc., vanish also in the zig-zag configuration. 

To characterise the entanglement between two sites two sets of each two conditions, as used in \cite{Anders08}, emerge. For each site $j$ these separability conditions are
\begin{eqnarray}\label{entcrit}
	0 \leq S_{1, 2}(\nu_t,\tau)=\frac{1}{\hbar^2}\left\langle (x_j \pm x_{j+\tau})^2\right\rangle
	\left\langle (p_{x_j} \mp p_{x_{j+\tau}})^2\right\rangle-1
\end{eqnarray}
and similarly for the $y$ direction. The expectation values needed here can be calculated using the transformation rules into the diagonal modes,  Eq.~(\ref{eq:FTlinear}) in the linear chain, and Eq.~(\ref{eq:trafo-zig-zag}) and Eq.~(\ref{eq:diamodes}) in the zig-zag configuration. If one of the inequalities is violated then entanglement exists between the $j$-th site and its $\tau$'s neighbour and the negativity,
\begin{equation}
	E_N=\sum_{k=1}^2 \max\left[0,-\ln \sqrt{S_k+1}\right],
\end{equation}
measures their degree of entanglement. The two criteria $S_1$ and $S_2$ witness two types of entanglement. 
For example, the EPR pair originally considered in \cite{EPR} shows violation for  $S_2$ but not $S_1$.

The von Neumann entropy of a single site $j$  in either $x$ or $y$ dimension is obtained following \cite{vNEntropy} with the formula
\begin{equation}
	S_V(r_j)=\frac{r_j+1}{2} \ln \left(\frac{r_j+1}{2}\right)-\frac{r_j-1}{2} \ln \left(\frac{r_j-1}{2} \right)
\end{equation}
where $r_j = \sqrt{\langle x_j^2 \rangle \langle p_{x_j}^2 \rangle }$, and similiarly for the $y$ direction, is the symplectic eigenvalue of its reduced state. To evaluate the entropy of a block of $n$ neighbouring sites the block entropy is then simply
\begin{equation}
	S_V(n)=\sum_{j=1}^n S_V(r_j)
\end{equation}
where the sum is taken over all $n$ symplectic eigenvalues $r_j$ in either $x$ or $y$ dimension within the block. 

\subsection{Thermodynamical limit ($N \to \infty$)} 
To obtain the negativity in $x$ direction at zero temperature in the thermodynamical limit in the linear configuration, we evaluate Eq.~(\ref{entcrit}) in the  ground state leading to 
\begin{equation*}
S_{1,2}(\nu_t,\tau)=\frac{1}{N^2}\sum_{l,k} \frac{\omega_{x,k}}{\omega_{x,l}}\left(1\pm \cos\left(\frac{2 \pi l}{N} \tau \right)\right) \left(1\mp \cos\left(\frac{2 \pi k}{N} \tau \right)\right)-1,
\end{equation*}
and similarly for the entanglement in $y$ direction. For the zig-zag configuration transformation \ref{eq:diamodes} gives a more complicated expression for the $S_{1,2}$ criteria. The formula is long and not enlightening, therefore we omit it.
In the thermodynamical limit we substitute $\frac{\pi k}{N}=\alpha $, $\frac{\pi l}{N}=\beta$, and replace the sums with integrals, 
\begin{equation*}
	S_{1,2}(\nu_t,\tau)=\frac{1}{\pi^2} \int_{-\pi/2}^{\pi/2} \omega_{x,\alpha} 
		\left(1\pm \cos\left(2 \alpha \tau \right)\right) d\alpha
		\int_{-\pi/2}^{\pi/2} \frac{\left(1\mp \cos\left(2 \beta \tau \right)\right)}{\omega_{x,\beta}}d\beta-1
\end{equation*}
These integrals can be evaluated numerically and the plots are shown in Fig.~\ref{fig:All} (red-solid line). For one special set of parameters the above expression can be easily calculated analytically, namely for $y$ direction in the linear configuration at the critical point $\nu_{t,crit}^2=\frac{C}{2}$ for NN interaction. The $S_1$ criterion between nearest neighbours becomes 
\begin{eqnarray}
	S_{1}(\nu_{t,crit},1)&=&\frac{4}{\pi^2} \int_{0}^{\pi/2} \sqrt{\frac{C}{2}}  \cdot \cos(\alpha) \left(1- \cos\left(2\alpha \right)\right) d\alpha \nonumber \\&\times & \int_{0}^{\pi/2} \frac{\left(1+ \cos\left(2\beta \right)\right)}{\sqrt{\frac{C}{2}} \cdot \cos(\beta) } d\beta-1=\frac{4}{\pi^2} \cdot \frac{2}{3} \cdot 2-1\approx-0.46
\end{eqnarray}
which leads to a value of the negativity of $E_N\approx 0.308$. 

In a similar fashion, the single-site von Neumann entropy for both, $x$ and $y$ direction, can be evaluated in the thermodynamical limit. Here we show the calculation for the $y$ direction. The symplectic eigenvalue for a  single site in $y$ direction is
 $r_j = \sqrt{\langle y_j^2 \rangle \langle p_{y_j}^2 \rangle }=\sqrt{\sum_{k,l=1}^N \frac{1}{N^2} \< Y_l^2\> \<P_{y_l}^2\>}$. Again we substitute sums with integrals and $\frac{\pi k}{N}=\alpha $, which gives an integral expression for the symplectic eigenvalue
  \begin{equation}
 r_j=\sqrt{\frac{1}{\pi^2} \int_0^{\pi/2} \frac{1}{\sqrt{\nu_t^2-C/2 \sin^2(\alpha)}}d\alpha \int_0^{\pi/2}\sqrt{\nu_t^2-C/2 \sin^2(\beta)} d\beta}.
 \end{equation}
 At the critical point $\nu_{t,crit}^2=\frac{C}{2}$ and for NN interaction this can be simplified to
 \begin{equation}
 r_j=\sqrt{\frac{1}{\pi^2} \int_0^{\pi/2} \frac{1}{\cos(\alpha)}d\alpha \int_0^{\pi/2}\cos(\beta) d\beta} \rightarrow \infty \hbox{ .}
 \end{equation}
The first integral diverges and hence the symplectic eigenvalue and the von Neumann entropy diverge at the QPT.

\section{Behaviour of entanglement at zero temperature} 

\begin{figure*}[t]
	\includegraphics[width=1.0\textwidth]{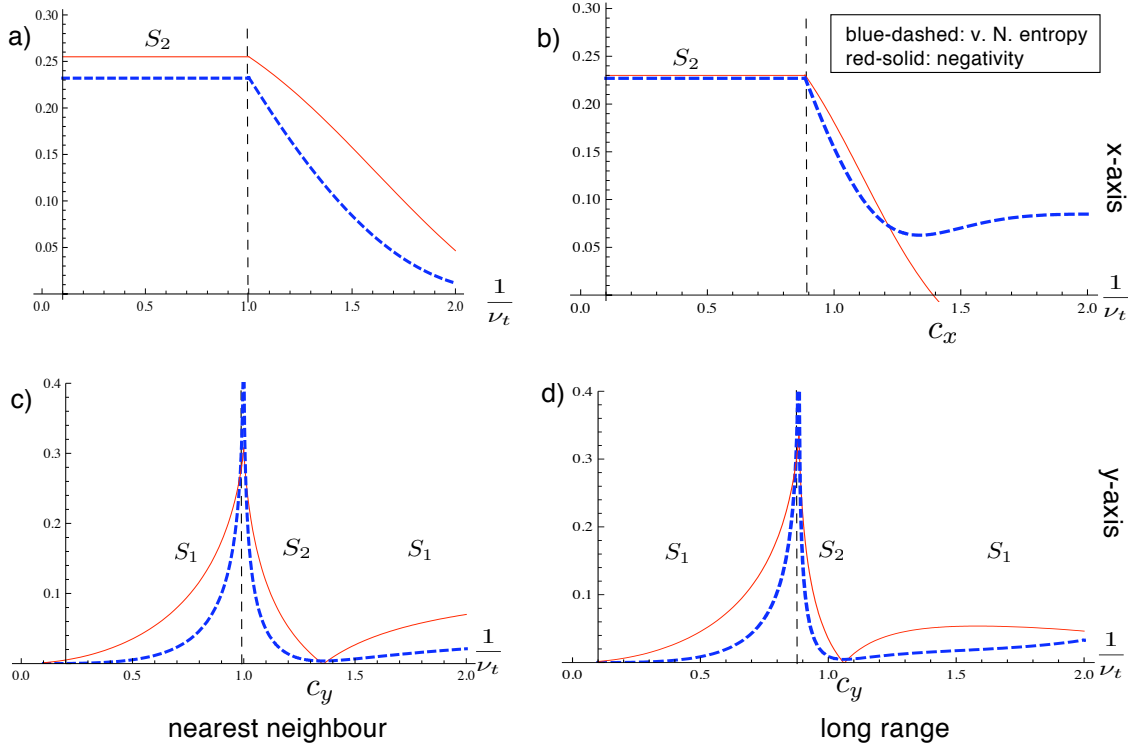}
	\caption{\label{fig:All} This graphic shows the von Neumann entropy of a single site and the negativity between two neighbouring sites both, for $x$-entanglement (upper plots) and $y$-entanglement (lower plots) of the ground state ($T=0$).  We compare the case where only nearest neighbour interact (left plots) with the long-range Coulomb Hamiltonian (right plots). Both models are approximated up to second order. The numerical values for the plots are for the ground state ($T=0K$) in the thermodynamical limit ($N\rightarrow \infty$) and $Q=1$, $m=2$. For NN the lattice constant is set to $a=1$, and for LR it is $a=14/15$ and the plots include interactions up to the forth neighbour. The change of entanglement at the critical point, indicated by the vertical line, is clearly visible in all four plots. }
	\end{figure*}

Fig.~\ref{fig:All} displays both entanglement measures for decreasing transverse trapping potential $\nu_t$. In the upper plots  (a and b) both measures for the $x$-entanglement are constant in the linear regime. This is  because the phonons in $x$ direction are independent of the trapping in $y$ direction. At the critical point both negativity and entropy are not differentiable. Decreasing the trapping potential beyond the critical value, where the zig-zag configuration is formed, the $x$-entanglement is reduced due to the emerging $x-y$ coupling. Because each site in (b) couples to several different neighbours, the entanglement between two distinct sites disappears. 

In the lower plots (c and d) both entanglement measures for the $y$-entanglement grow with decreasing trapping potential $\nu_t$ in the linear configuration. The even numbered ions oscillate exactly out of phase with the odd numbered ions due to the repulsive Coulomb potential. The smaller $\nu_t$, the larger these quantum fluctuations around equilibrium position become. At the critical point the fluctuations become strong enough for causing the ions to move outward. The entropy diverges and the negativity reaches its maximal  value of $E_N\approx 0.308$ where it is not differentiable. 

For the nearest neighbour coupling (a and c) the negativity and entropy show qualitatively the same behaviour. This can be understood easily as any entanglement of an site with the rest of the chain, measured by the von Neumann entropy, is created by the coupling with only the nearest neighbours. For the long-range interaction (b and d), where a single site couples to all other sites, there are significant differences between the two entanglement measures. While the negativity of two nearest neighbour sites vanishes after a threshold value of $\nu_t$, each single site remains entangled with the rest of the chain, as seen by the positive value of the von Neumann entropy.

Both entanglement measures are functions of the eigenvalues of the Hamiltonian. Therefore abrupt changes in entanglement can signal non-analyticity of ground state energy, which is associated with QPT's. In \cite{Lidar} it was shown for finite dimensional systems that (under certain conditions)  a discontinuity in or divergence of the first derivative of the negativity is both necessary and sufficient to signal a QPT. It seems intuitive that a similar characterisation holds also for continuous variable systems. 
Here the divergence of entropy and finite discontinuity of the first derivative of negativity shows a QPT of second order. After the phase transition the violated negativity criterion switches from $S_1$ to $S_2$. Lowering the trapping potential further leads to another ``critical'' point $c_y$ at which the negativity becomes zero while the entropy reaches its minimal (non-zero) value. 

\begin{figure}[t]
	\begin{center}
	\includegraphics[width=0.5\textwidth]{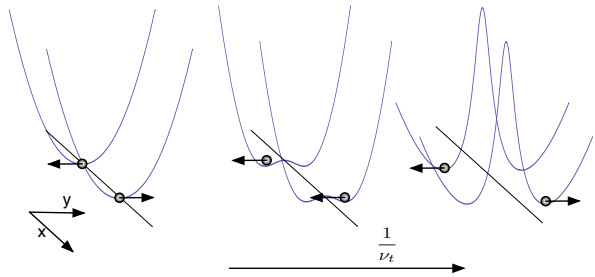}
	\caption{\label{fig:potentials} This graphic shows the change of trapping potential in $y$ direction. Different geometries favour different momenta, as indicated with the arrows.}
	\end{center}
\end{figure}

The additional critical point $c_y$ (and $c_x$ in (b)) only appears when the interaction is harmonic as is the case in our second order approximation of the Coulomb model. The point is due to a sign change of the second order coefficient $d^y_{\tau}$ ($d^x_{\tau}$) at $c_y$ ($c_x$). As a consequence the interaction switches from repulsive to attractive. When anharmonic terms are taken into account, as in the numerical treatment in \cite{Plenio},  these points vanish. An intuitive way of understanding the switching between $S_1$ and $S_2$  is as follows, see Fig. \ref{fig:potentials}. Decreasing the trapping potential changes the relative strengths of the inner and outer potential for the motion in the $y$ direction. This leads to the change in phase in the relative momenta of neighbouring sites. In one configuration the relative potential favours momenta in the opposite directions, while the other configuration favours motion in the same direction. This is reflected in the change from $S_1$ to $S_2$. However, although the negativity vanishes at these points, a single site is still entangled with the rest of the chain.  

\subsection{Block Entropy}

\begin{figure}[t]
	\begin{center}
	\includegraphics[width=\textwidth]{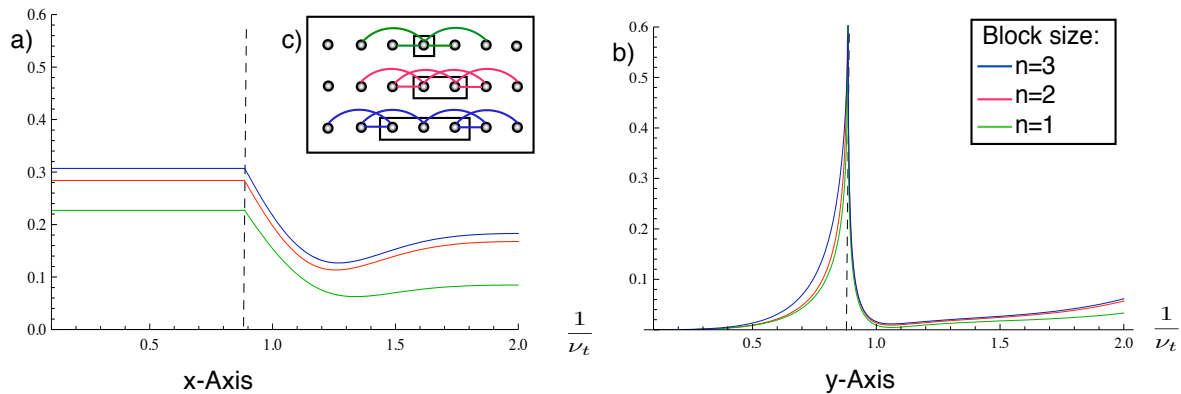}
	\caption{\label{fig:blockentropy} This graphic shows the block entropy for different number of sites in $x$ (a) and $y$ (b) dimension for long range interaction in the thermodynamical limit. The entropy increases with number of sites. The inset (c) shows the entanglement connections between nearest neighbour and next nearest neigbour (NNN) sites.}
	\end{center}
\end{figure}

The block entropy measures how much entanglement exists between a block of sites of the lattice and the rest. For nearest neighbour interaction models there exist scaling laws showing that the amount of entanglement scales with the boundary area and not the volume of the reduced state \cite{ReviewAreaLaw}. For a translational invariant chain with NN interaction there exists even a computable analytical result for the negativity of a bisected harmonic chain \cite{Audenaert}. Here we investigate the block entropy for blocks up to three sites in the long-ranged Coulomb lattice. Due to the complexity of LR interactions few results are known so far. The inset (c) illustrates the increase of correlation across the block boundary with increasing block size. Correlations stretch to nearest neighbours (NN) and next-nearest neighbour (NNN), as shown in Fig. \ref{fig:blockentropy} while third neighbour entanglement is negligible. Increasing the block size from one to two there are twice as many NNN connections across the boundary and hence the block entropy for two sites is expected to increase. This intuition is confirmed in a) and b) showing the block entropy in $x$ and $y$ direction, repectively. However, it can be seen that the while the entropy increases slightly with number of sites, no qualitative difference can be observed.  This is due to the fact that the additional, long range entanglement is much weaker because the Coulomb potential falls of quickly, with $1/\tau^3$ where $\tau$ is the distance of sites.  In $y$ dimension (b) already the single site entropy is a good approximation for larger blocks. This is because transversal next nearest neighbour entanglement turns out to be very small. In $x$ direction (a) there are significant differences for growing block sites. Here next nearest neighbour entanglement cannot be neglected.  Although the number of connections between two and three sites is the same, the block three entropy is still higher. This might be due to multipartite entanglement. Larger block sites are difficult to evaluate, as the symplectic eigenvalues of the covariance matrix become very complicated.

\section{Witnessing entanglement at finite temperature} 

One consequence of the entanglement of the sites is a lowering of the energy of the system \cite{Anders08}. This can be seen by assuming that the thermal state of the system
is separable, i.e. decoupled between modes. Then an effective, single site  Hamiltonian can be obtained by removing all second order couplings between the different sites, i.e. $\langle (x_{j}-x_{j+\tau})^2 \rangle_{\rho_S}=\langle x_j^2+x_{j+\tau}^2-2 x_j x_{j+\tau} \rangle_{\rho_S} = 2 \langle x_j^2 \rangle_{\rho_S}$ etc.. The total Hamiltonian becomes the sum of the single site Hamiltonian $H_{\rm eff}=\sum_j H_j$ with
\begin{equation}
	H_j = \frac{p_{xj}^2+p_{yj}^2}{2m} +\frac{m}{2}\left(\Omega_x^2 x_j^2+\Omega_y^2 y_j^2+ \Omega_{xy} x_j y_j \right) 
\end{equation}
where
 \begin{eqnarray}
\Omega_x =\sqrt{\nu^2+\frac{2 Q^2}{m}\sum_{\tau \neq 0} d_\tau^x} 
\, \mbox{ and } \,
\Omega_y =\sqrt{\nu_t^2+\frac{2 Q^2}{m}\sum_{\tau \neq 0} d_\tau^y} \nonumber \\
 \mbox{ and } \, \Omega_{xy}=\frac{2 Q^2}{m}\sum_{\tau \neq 0} d_{\tau,j}^{xy}
\end{eqnarray}
Then the thermal state takes the form
\begin{equation}
	\rho_S=\bigotimes_{j=1}^N \frac{e^{-\beta H_j}}{\tr\left[e^{-\beta H_j}\right]},
\end{equation}
with $\beta=\frac{1}{k_B T}$ the inverse temperature.
Using the transformations  Eq.~(\ref{eq:FTlinear}), Eq.~(\ref{eq:trafo-zig-zag}) and Eq.~(\ref{eq:diamodes}), $\< H_{\rm eff} \>$ can be fully diagonalised. The internal energy $U=\<H_{\rm eff}\>$ for any separable state is bounded from below by zero point fluctuations, i.e.
\begin{equation}\label{entWitness}
	\<H_{\rm eff}\>_{sep}\geq \frac{N \hbar}{2} \left(\Omega_{x}+\Omega_{y}+\Omega_{xy} \right),
\end{equation}
and any state having a smaller energy must be entangled between the individual sites.

\begin{figure}[t]
	\begin{center}
	\includegraphics[width=0.7\textwidth]{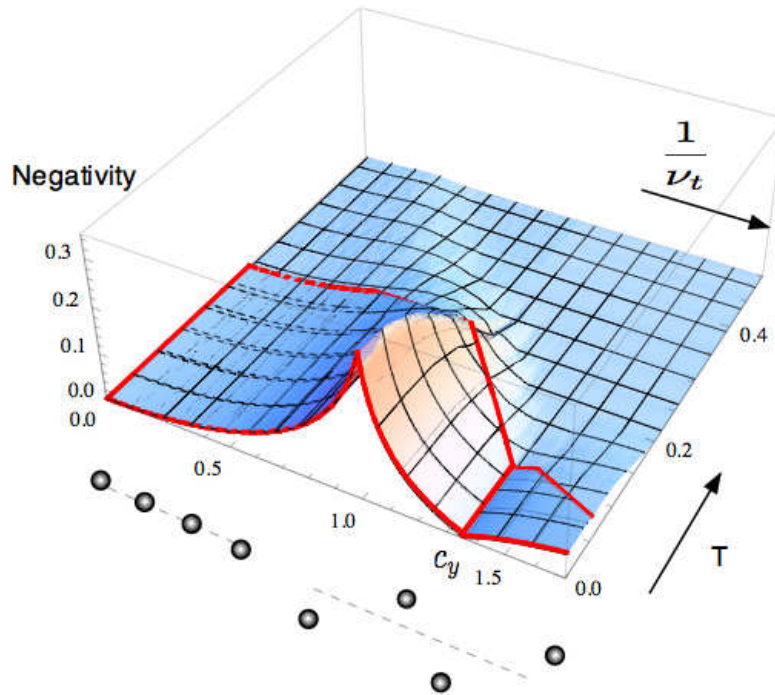}
	\caption{\label{fig:Tempy} This graphic shows the negativity in $y$ direction of two neighbouring ions for decreasing trapping potential and increasing temperature (in units of $[\nu_t]=\sqrt{\frac{Q^2}{m a^3}}$ and $[T]=[\nu_t] \frac{\hbar}{2 k_B}$, $N = 20, Q = a= 1, m=2$). The red line indicates the critical temperature, obtained with the energy witness argument, below which entanglement of some form must be present in the chain. }
	\end{center}
\end{figure}

We now want to see how the ground state situation is modified at non-zero temperature. As the energy of the thermal state, i.e. the mean excitation of phonons, increases with temperature there exists a critical temperature, $T_c$, for each value of the trapping potential at which the thermal state matches the energy bound. The negativity between the $y$ degrees of freedom of two nearest neighbours for NN Hamiltonian is evaluated numerically and plotted in Fig.~\ref{fig:Tempy} where the critical temperature for full separability is also indicated as a red-line. When the trapping potential is lowered, the negativity increases until reaching its maximal value at the critical $\nu_{t, crit}$. Further lowering leads to a decrease of the negativity until it vanishes at point $c_y$. However, when further decreasing $\nu_t$ the negativity grows again, yet the two entanglement criteria $S_{1,2}$ are now switched. As expected, increasing the temperature leads to smaller values of negativity and smoothens the entanglement measures to make it differentiable at the critical point.  For large $\nu_t$ the negativity is small, but remains finite until relatively high temperature. The sharp peak at the QPT remains almost constant for finite $T$ and decreases fast. Thermal states within the red outlined area have a smaller energy than any separable state and their entanglement is therefore detected by the energy witness. Remarkably, states at $c_y$ are entangled for temperatures up to $T_c [1/\nu_t=c_y]=0.12\frac{ [\nu_t] \hbar}{2 k_B}$, even though there is no nearest neighbour entanglement in $y$ direction. This is because there is still nearest neighbour entanglement in $x$ direction and possibly also multipartite entanglement the chain.

\section{Conclusions}\label{sec:conc}

In this paper, we revised the classical phase transition in a long range harmonic chain \cite{Classical} using a fully quantised model. Two measures of entanglement display critical behaviour: The von Neumann entropy of  a single site and blocks of two and three sites diverge at the critical point while the negativity  is not differentiable. Thus also in this continuous variable  system entanglement indicates a QPT, as previously shown for discrete systems \cite{Lidar}. 
The negativity depends only on single site and nearest neighbour correlations; the single site von Neumann entropy even only depends on single site measurements. Our calculation shows that even this local, single site function is able to detect a global change in configuration. This implies that instead of examining two point correlations functions one can alternatively consider the entanglement measures stated. This will be advantageous in experimental situations when the number of different measurement procedures is best kept as low as possible. We are aware that for the moment ion traps cannot yet perform the required measurements of e.g. single site variance of space and momentum operator, but this is a technical, not a fundamental problem.
Furthermore, our results confirm that this phase transition is of second order as indicated in \cite{Classical}. At finite temperature, the negativity still displays critical behaviour, as seen in Fig.~\ref{fig:Tempy}, however the non-differentiable cusp fades out quickly with increasing thermal noise. Tuning across a QPT provides a means of changing the amount and structure of continuous variable. Experiments with ion-traps are ideally suited to study QPTs with great precision. 
 
\ack The authors wish to thank Dieter Jaksch for inspiring discussions. E.R. is supported by the National Research Foundation and Ministry of Education, in Singapore. J.A. is grateful for support by the EPSRC's QIPIRC program in the UK and thanks the Centre for Quantum Technologies in Singapore for hospitable stays. V.V. acknowledges financial support from the Engineering and Physical Sciences Research Council, the Royal Society and the Wolfson Trust in UK as well as the National Research Foundation and Ministry of Education, in Singapore. 

\vspace{2cm}

\noappendix

\end{document}